\documentclass[amsmath,
 amssymb,
 aps,
 nofootinbib,
 preprintnumbers
]{revtex4-1}
\usepackage[utf8]{inputenc}
\usepackage{simplewick}
\usepackage{footnote}
\usepackage{graphicx}
\usepackage{dcolumn}
\usepackage{bm}
\usepackage{stmaryrd}
\usepackage{amsmath}
\usepackage{amssymb}
\usepackage{bbm}
\usepackage{amsfonts}
\usepackage{xcolor}
\usepackage{footmisc}
\usepackage{comment}
\usepackage{ulem}
\usepackage{slashed}
\usepackage{multirow}
\usepackage{appendix}
\usepackage{float}
\usepackage[linktocpage=true,bookmarksnumbered=true,colorlinks=true,plainpages,linkcolor=blue,citecolor=red]{hyperref}

\begin{document}

\title{Constraining dark boson decay using neutron stars\footnote{Invited Review Article for a Special Edition on Neutron decay anomalies for \texttt{UNIVERSE}.}}

\author{Wasif Husain, Dipan Sengupta  and A W Thomas} 

\affiliation{ \quad ARC Centre of Excellence for Dark Matter Particle Physics, Department of Physics, The University of Adelaide, SA 5005, Australia.}
\email{wasif.husain@adelade.edu.au,dipan.sengupta@adelaide.edu.au,anthony.thomas@adelaide.edu.au}
\begin{abstract}
    Inspired by the well known anomaly in the lifetime of the neutron, we investigate its consequences inside neutron stars. We first assess the viability of the neutron decay hypothesis suggested by Fornal and Grinstein within neutron stars, in terms of the equation of state and compatibility with observed properties. This is followed by an investigation of the constraint information on neutron star cooling can place on the decay rate of the dark boson into standard model particles, in the context of various BSM ideas.
\end{abstract}

\maketitle
\section{Introduction}
\label{sec:intro} 
Neutrons are a fundamental constituent of our universe. It has been over almost a century since they were discovered but their lifetime ~\cite{RevModPhys.83.1173} still presents a challenging problem to solve. In particular, current experiments appear to show a difference in the neutron lifetime when measured with different methods. In the bottle method~\cite{Serebrov:2017bzo,Pattie:2017vsj,TAN2019134921,PhysRevC.85.065503} the neutrons are trapped and the number counted after a fixed time, with no specific determination of the decay mode. In contrast, using beam 
method~\cite{PhysRevLett.111.222501,Otono:2016fsv,Olive_2016} one actually observes the protons produced in $\beta$ decay. Of course, the lifetime of the neutron should be the same, regardless of the method of measurement. However, the lifetime of the neutron shows a discrepancy. Using the bottle method,  Ref.~\cite{UCNt:2021pcg} found the lifetime of the neutron to be 877.75 $\pm 0.28_{\rm stat} + 0.22 - 0.16_{\rm syst}$ s, which is very close to the lifetime measured in Ref.~\cite{doi:10.1126/science.aan8895} and Ref.~\cite{PhysRevC.97.055503}. On the other hand, using the beam method the lifetime measured has been measured to be 
887.7 $\pm 0.7 \, (stat) \, +0.4/-0.2 \, (sys)$ s~\cite{doi:10.1126/science.aan8895}. 

A resolution of this discrepancy in the lifetime of the neutrons could potentially lead to new physics. A solution along those lines was recently proposed by Fornal and 
Grinstein~\cite{Fornal:2018eol,Grinstein:2018ptl,Fornal:2020gto}, who proposed an extra decay channel of the neutron into dark matter. Based on the difference in lifetimes, they suggested that roughly 1\% of the time neutrons decay into dark matter, wile for the remaining 99\% of the time they undergo $\beta$ decay. In the beam method the dark matter would go undetected and uncounted, while in the bottle method the effect of dark matter is automatically included.  

According to the hypothesis of Fornal and Grinstein the dark decay mode of the neutron is
\begin{equation}
\label{FnG}
    n \longrightarrow \chi + \phi \, ,
\end{equation}
where $\chi$ is the dark fermion and $\phi$ the dark boson. This hypothesis has attracted the interest of many physicists. For example, experimental studies showed very quickly that the $\phi$ particle could not be a photon~\cite{Tang:2018eln,Serebrov:2007gw}. The hypothesis was also very rapidly subject to tests using the properties the neutron stars, which indicated that the dark fermions, $\chi$, has to experience a strong vector repulsion in order to be consistent with the observations~\cite{Motta:2018rxp,Motta:2018bil,PhysRevLett.121.061801,2018_sa}. A recent study~\cite{Husain:2022bxl} suggested that there might be an observable signal of this decay if one could observe the neutron star right after its birth. An interesting discussion about the neutron decay can be found in ~\cite{Ivanov:2018vit}.

An alternative to the Fornal and Grinstein hypothesis was proposed by Strumia in Ref.~\cite{Strumia:2021ybk}, where the author suggested that the neutrons might decay into three identical dark fermions, $\chi$, 
\begin{equation}
\label{Stru}
    n \longrightarrow \chi + \chi + \chi, 
\end{equation}
with each of them having baryon number 1/3 and mass $m_\chi$ = (mass of neutron)/3. An advantage of this proposal was that, to be consistent with the constraint on the maximum mass of neutron stars, the dark fermions, $\chi$, are not required to be self-interacting. Our recent study in Ref.~\cite{Husain:2022brl} on the Strumia hypothesis agrees with this claim and indicated that one could find observable signals of neutron decay similar to those found within the Fornal and Grinstein hypothesis.

Within the Fornal and Grinstein hypothesis, shown in 
Eq.~(\ref{FnG}), the mass of the decay products must be in the range 937.9 MeV <  $m_\chi +m_\phi$  < 938.7 MeV for the known stable nuclei to remain stable~\cite{Fornal:2018eol,Grinstein:2018ptl}. To date, most of the studies on the hypothesis using neutron stars, have considered the $\phi$ as an extremely light particle which escapes the neutron star immediately and treating the $\chi$ as almost degenerate with the neutron. 
There is a possibility that if $\phi$ boson has a mass close to the difference of the masses of the neutron and the $\chi$, for example of order 1 MeV, then it may remain trapped inside the neutron star. 

Here, the focus is on trapped $\phi$ bosons and their effects on neutron star heating. This leads to a strong constraint on the lifetime of the $\phi$, which is then compared with limits from other studies of dark matter candidates. The manuscript is divided into sections as follows. Section~\ref{NS model} covers the necessary model for the equation of state of nuclear matter inside the neutron star and explains the change associated with neutron decay into $\chi$ and $\phi$. This is followed by 
section~\ref{results}, where the consequence of trapping the  $\phi$ boson are explored. In section~\ref{bosonDecay} the decay modes of the $\phi$ boson have been studied in detail. Finally, section~\ref{conclusion} presents a summary of our findings.
\section{Neutron stars}
\label{NS model}
Neutron stars are comparatively small objects that come into existence when an ordinary star of mass 8 M$_\odot$ - 15 M$_\odot$ dies. Neutrons are not surprisingly the dominant component of a  neutron star and if neutrons decay into $\chi$ and $\phi$ then this decay must also take place inside the neutron star. Therefore, the neutron stars must contain $\chi$ fermions and in the circumstances explained earlier also $\phi$ bosons, and their presence inside neutron stars must change their properties  \cite{Mukhopadhyay_2017,PhysRevD.77.043515,PhysRevD.77.023006,CIARCELLUTI201119,Sandin_2009,Leung:2011zz,Ellis_2018,bell2020nucleon,2021_w,2000NuPhB.564..185M,Blinnikov:1983gh,2018_sa,2019_reddy,2013_red,berryman2022neutron,McKeen:2021jbh,deLavallaz:2010wp,Busoni:2021zoe,Sen:2021wev,Guha:2021njn}. There are some strong constraints on the properties of neutron stars imposed by observations that a realistic neutron star model must follow. For example, Ref.~\cite{LIGOScientific:2018cki} showed that a neutron star of mass 1.4 M$_\odot$ should have a radius 10 - 14 km. 
PSR J1614-2230 \cite{2010Natur.467.1081D} and PSR J0348+0432 \cite{2013Sci...340..448A} have masses 1.928 and 2.01 M$_\odot$, respectively, so the neutron star model must predict maximum mass of neutron stars at least 2 M$_\odot$ \footnote{Note that recent observations of pulsars PSR J0030+0451\cite{Riley:2019yda}, and PSR J0740+6620 \cite{Riley:2021pdl} constrain their masses and radii to be 1.34 M$_{\odot}$ and 12.71 km and 2.072  M$_{\odot}$ and 12.39 km respectively. See also \cite{Miller:2019cac,Miller:2021qha}. }. The tidal deformability should be consistent with the 
discovery~\cite{Abbott_2017,Abbott_2019} of gravitational wave detection by LIGO and VIRGO observatories.
Therefore neutrons stars can be very helpful in testing the Fornal and Grinstein hypothesis.  

Neutron star interiors covers a wide range of densities right from the surface to the core ~\cite{Lawley:2006ps,Whittenbury:2013wma,Whittenbury:2015ziz,PhysRevD.4.1601,PhysRevD.30.272,Bombaci_2004,PhysRevD.102.083003,doi:10.1143/JPSJ.58.3555,2012_a,doi:10.1063/1.4909561,2016_a,PhysRevC.58.1804,BALBERG1997435,1985ApJ...293..470G,KAPLAN198657,PhysRevLett.79.1603,PhysRevLett.67.2414,Glendenning1997,Haensel2017,1980PhR....61...71S,weber2007neutron,2019_fri,Weber2016,Terazawa:2001gg,Husain_2021,2001_lattimer,2020_latti,2021_l,Cierniak:2021knt,Shahrbaf:2022upc,10.1143/PTP.108.703,2017_xyz,2022_xxyz,Motta:2022nlj}, with the cores containing the most dense matter in the universe. To model the neutron star, one needs to adopt a suitable equation of state for the nuclear matter. At the core of a neutron star the density could be as high as 6 times the density of normal matter. Therefore, one needs to chose a model capable of describing the physics at such high densities. In this study quark meson coupling (QMC) model~\cite{Guichon:1987jp,Guichon:1995ue,Stone:2016qmi,RIKOVSKASTONE2007341} is adopted to model the neutron star matter. A brief description of the QMC model is presented below. 

\subsection{Quark Meson Coupling model}
The quark meson coupling model was initially proposed by Guichon~\cite{Guichon:1987jp} and further developed by Guichon, Thomas and collaborators~\cite{Guichon:1995ue,Saito:2005rv}. In this model the nucleons are treated as a collection of 3 quarks confined in an MIT bag~\cite{DeGrand:1975cf}. The internal structure of the nucleon is treated with great importance, unlike other models where nucleons are considered as a point like objects. In the QMC model the interaction between baryons is generated by the exchange of mesons, which couple self-consistently to the confined quarks. The strong scalar mean field in particular drives significant changes in the structure of the bound baryons.  

The equation of state based on the QMC model has shown been shown to lead to an acceptable description of neutron star properties~\cite{RIKOVSKASTONE2007341,Motta:2022nlj,Husain:2022brl}. The effective mass of the nucleon in-medium may be expressed in terms of the scalar polarisability, '$d$', the mass of the free nucleon, $M_N$, and the coupling constant of the $\sigma$ field to the nucleon in free space, $g_\sigma$, as
\begin{equation}
M_N^*(\sigma) = M_N -  g_\sigma \sigma + 
\frac{d(g_\sigma \sigma)^2}{2}  . 
\label{eq2.1}
\end{equation}
The details of QMC model can be found in Refs.~\cite{Guichon:1987jp,Guichon:1995ue,Guichon:2018uew}.
For simplicity, in this study it is assumed that neutron stars do not contain hyperons or strange matter at the higher energy densities, a nucleon only equation of state is 
used~\cite{Husain:2022bxl}. 

\subsection{Formalism including neutron decay}
According to the hypothesis given in Eq.~(\ref{FnG}), the neutron stars must contain $\chi$ and $\phi$. As mentioned above there is no a priori constraint on the mass of the $\phi$, within the small window allowed. We have chosen to study the case $m_\phi$ = 1 MeV, $m_\chi$ = 937.7 MeV in this work, where $m_\phi$ is the mass of $\phi$ boson and $m_\chi$ the mass of the $\chi$ fermion. 
In this case the velocity of the $\phi$ boson is sufficiently low that it will be trapped inside the neutron star. 

Inside the neutron star the  $\phi$ bosons must condense, according the Bose-Einstein condensation theory~\cite{Husain:2022bxl,Motta:2019tjc}. But the total contribution of the $\phi$ will be far too small to make any significant changes in the existing mass, radius, and tidal deformability constraints, even after condensation. In fact, we will see in later sections that the contribution of the $\phi$ bosons to the total mass is only about 1/10$^6$ M$_\odot$. Although this is small, nevertheless amount of mass may contribute to the heating of the neutron star if the $\phi$ bosons decay into standard model particles. Therefore the focus of this study is on the $\phi$ boson decay into standard model particles inside neutron stars.

The equations are solved using Hartree-Fock approximation. The full Hartree-Fock terms can be found in  \cite{Husain:2022bxl,Motta:2019tjc,Krein:1998vc}. 
Although the $\phi$ bosons will be in the lowest possible quantum state, the dark fermions will constitute a gas of fermions. The dark fermions are assumed to be self-interacting, in order to survive against the observational constraints on the neutron star properties. The self-interaction of the dark fermions is assumed to be similar to the neutron-$\omega$ interaction. 

The presence of $\chi$ and $\phi$ inside the neutron stars changes their composition. Therefore, the chemical equilibrium equations are~\cite{Motta:2018bil,Motta:2018rxp}
\begin{equation}
    \mu_n = \mu_\chi + m_\phi \qquad
 \mu_n = \mu_p + \mu_e  \qquad
 \mu_\mu = \mu_e  \qquad
 n_p = n_e + n_\mu  \, \qquad
\end{equation}
where $\mu$ represents the chemical potential of the different associated particles, and $n_p$, $n_p$ and $n_e$ stand for the number of neutrons, protons, electrons. The dark fermions, dark bosons and nuclear matter particles are assumed to not interact with each other. Therefore, the neutron star contains two non-interacting fluids. However, because the contribution of the dark matter compared to nuclear matter is very small, using the two-fluid TOV equation is not necessary. Therefore for ease of the calculation one fluid TOV is used.

\subsection{Tolman Oppenheimer Volkoff (TOV) equations}
To calculate the properties of the neutron star the equation of state is combined with the structural equations derived using the Einstein's equations of general relativity. Therefore, the TOV equations ~\cite{Tolman169,PhysRev.55.374,CIARCELLUTI201119,Sandin_2009} given as
\begin{equation}
    \frac{dP}{dr} = - \frac{[\epsilon(r)+P(r)][4\pi r^3(P(r) +  P(r))+m(r)]}{r^2(1-\frac{2m(r)}{r})},
\end{equation}
\begin{equation}
    m(r) = 4\pi\int_{0}^{r}dr.r^2 \epsilon(r) ,
\end{equation}
are integrated from the centre of the neutron star towards the surface, using the boundary conditions that at the surface the pressure and energy density should be zero. Here, $P$ is the total pressure, $P = P_{nucl} + P_{DM}$  and $\epsilon = \epsilon_{nucl} + \epsilon_{DM}$ is the total energy density, including the energy density of nuclear matter and dark matter.  
The tidal deformability is calculated by using the method explained in Ref.~\cite{Hinderer_2008,Hinderer_2010}.

\section{Results}
\label{results}
In this section the consequences of the neutron decay on the properties of the neutron stars are given. The vector interaction of dark fermions is increased until it follows the constraint ~\cite{Motta:2018bil,Grinstein:2018ptl,PhysRevLett.121.061801,Cline:2018ami,Berryman:2022zic,Strumia:2021ybk,2021_ramani,2018_tanga,2021_osc} on the properties of the neutron stars. 

\begin{figure}[ht]
\centering 
    \includegraphics[width=1\textwidth]{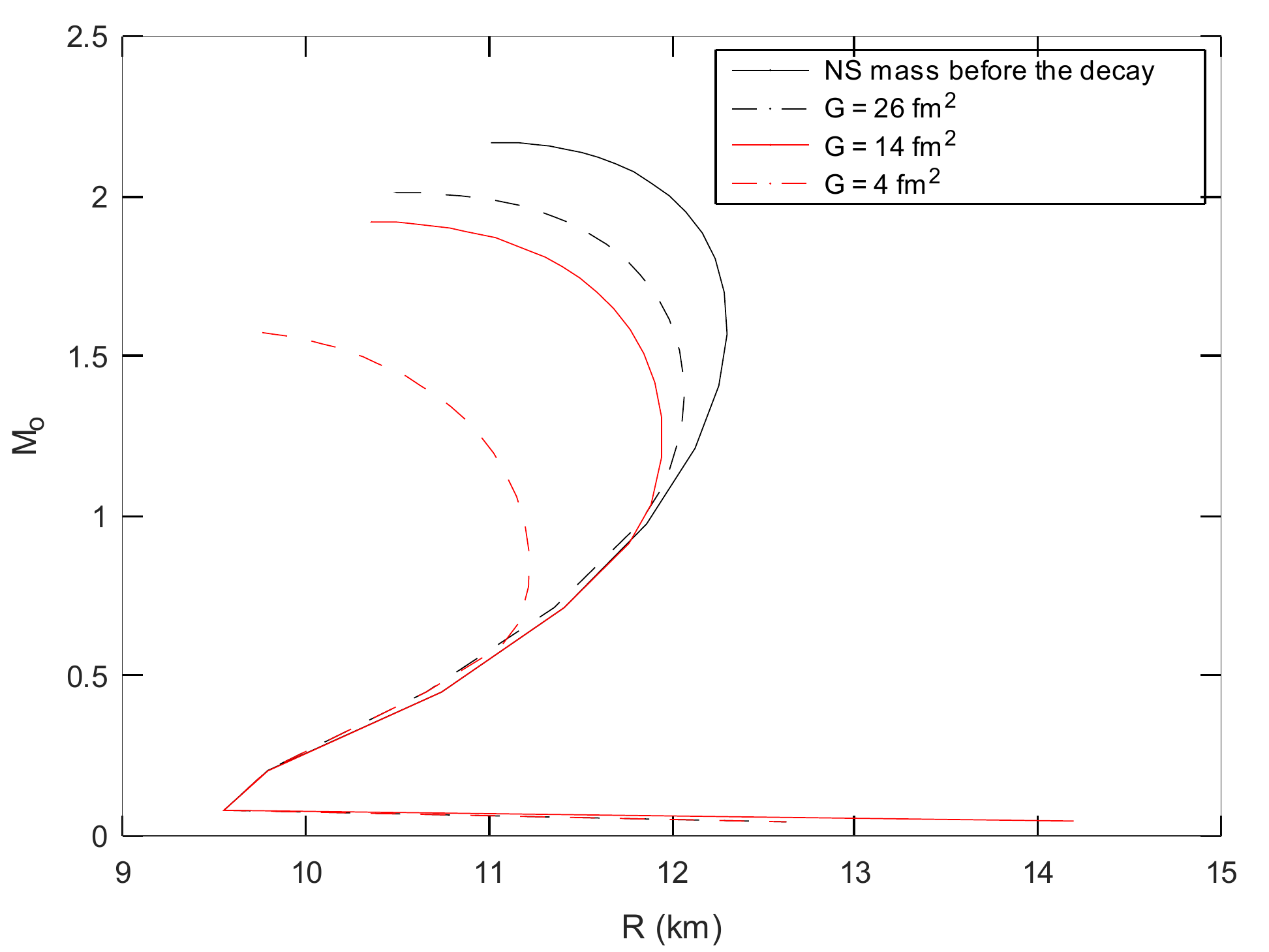}
\caption{The total mass (given in solar masses) vs radius of the neutron star. Here $G$ is the $\chi$-$\chi$ vector repulsion strength which is increased to satisfy the observational constraints.}
\label{fig1}
\end{figure}

As shown in Fig.~(\ref{fig1}), the mass of the neutron star is reduced after the neutron decay. In fact, the maximum mass of the neutron star falls below 2 M$_\odot$ after the neutron decay ~\cite{Motta:2018rxp,PhysRevLett.121.061801,McKeen:2018xwc} if the dark fermions are considered to be non-self-interacting. However, neutron stars of mass above 2 M$_\odot$~\cite{Ozel:2016oaf,2010Natur.467.1081D,2013Sci...340..448A} have been observed. Therefore, in order to survive, a neutron star model must predict neutron stars of maximum mass of at least 2 M$_\odot$. Fig.~(\ref{fig1}) indicates that the dark fermions must have self-repulsion with a strength parameter of order 26 fm$^2$ to be consistent with the observations. Moreover, there is a significant reduction in the radius of the neutron stars after the decay, which suggest that neutron stars should spin up during the decay. 

\begin{figure}[ht]
\centering 
    \includegraphics[width=1\textwidth]{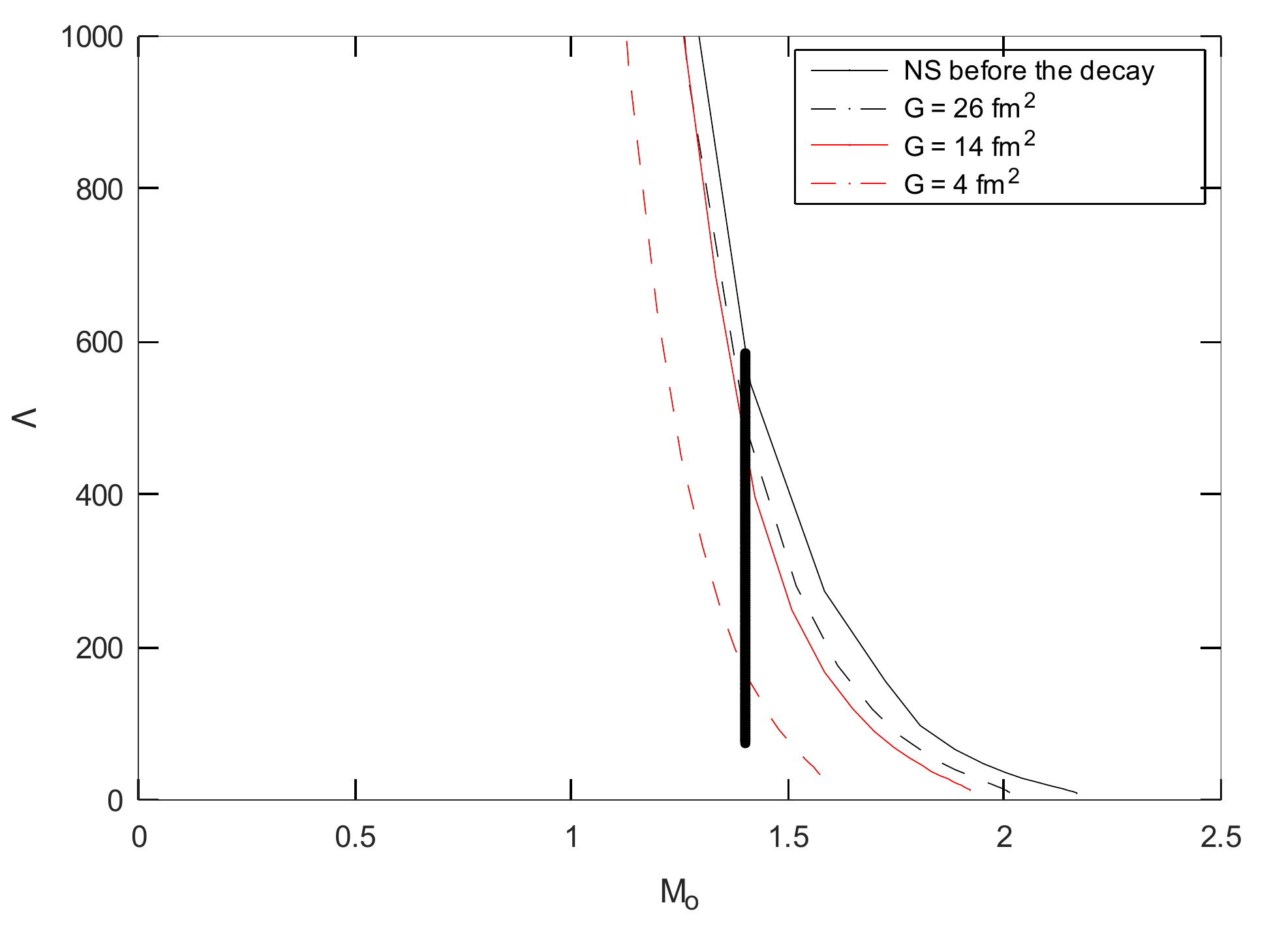}
\caption{Tidal deformability against the mass of the neutron star. Here, $G$ is the vector repulsion strength of dark fermions. The bold, black, vertical line indicates the acceptable range of values for tidal deformability \cite{Riley:2021pdl,Miller:2021qha}.}
\label{fig2}
\end{figure}

Figure~(\ref{fig2}) shows the tidal deformability against the radius of the neutron star. The analysis of the gravitational waves~\cite{Abbott_2017,Abbott_2019,Bramante_2018_31} indicated that a neutron star of mass 1.4 M$_\odot$ must have tidal deformability in the range 70 - 580, with 90\% confidence level. Fig.~(\ref{fig2}) shows that dark fermions with a vector self-repulsion of strength 26 fm$^2$ satisfy the constraint on mass and tidal deformability. 

\begin{figure}[ht]
\centering 
    \includegraphics[width=1\textwidth]{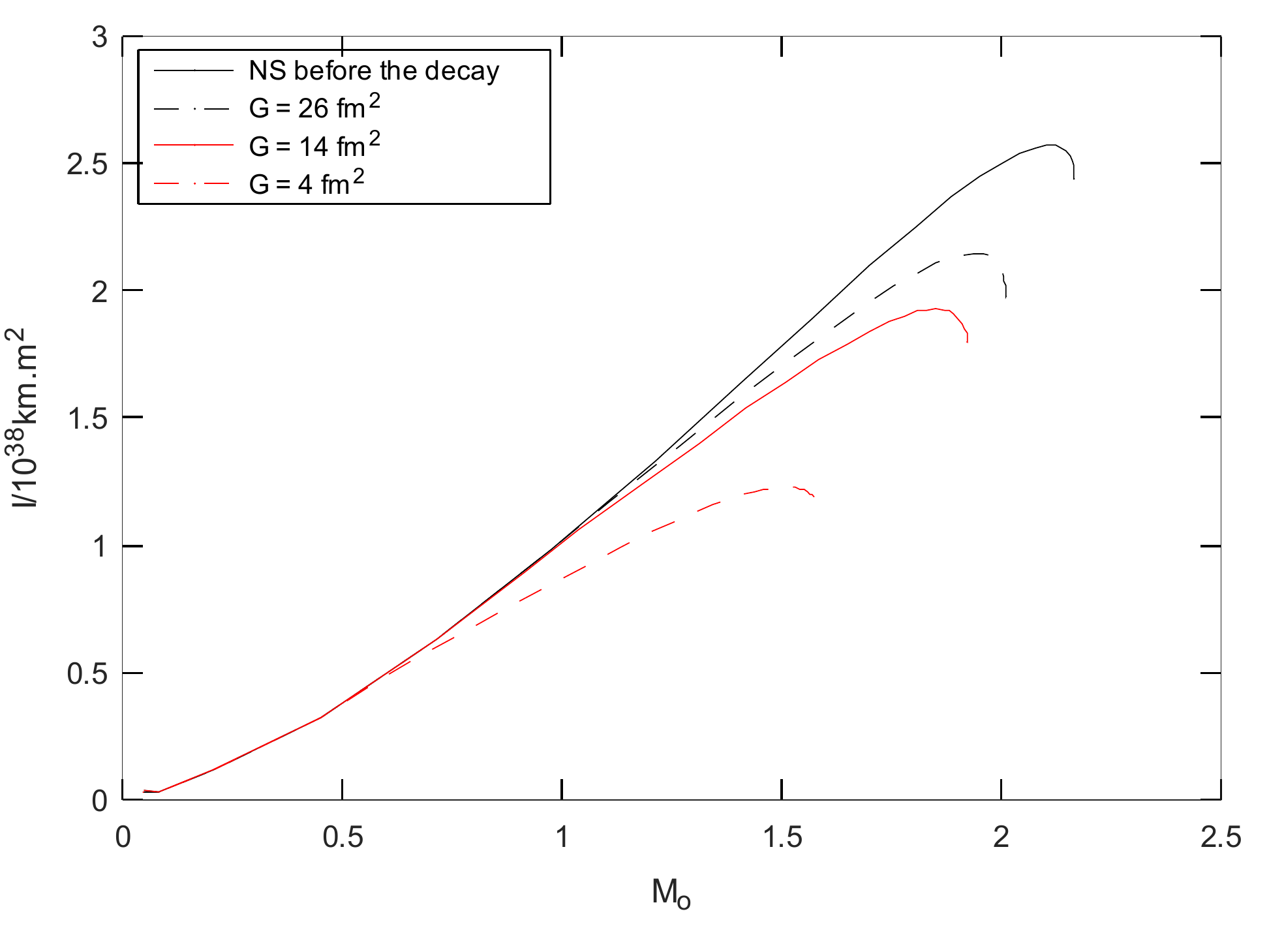}
\caption{Total mass versus the moment of inertia of the neutron star at with different self-interaction strengths of $\chi$s.}
\label{fig3}
\end{figure}

Figure~\ref{fig3} shows the moment of inertia against the mass of the neutron star. The moment of inertia is reduced after the neutron decay. When the dark fermion vector-interaction is lowered the difference in moment of inertia increases. Thus we are only interested in the case when the dark fermions, $\chi$, have a vector interaction strength $\ge$ 26 fm$^2$. 
Figure~\ref{fig3} indicates that the moment of inertia of heavier neutron stars is significantly reduced even when $G \ge$ 26fm$^2$, which should result in spinning up of the neutron star. That in turn may provide a signal of the neutron's exotic decay.
%

Most studies indicate that neutron stars cool down very quickly by the standard Urca process. After approximately a million years the neutron stars have a luminosity of order 10$^{31.5}$ erg/s. Therefore, if the $\phi$ boson decays into photons it will contribute to the heating of the neutron star and after a million years it must not contribute a luminosity $\ge$ 10$^{31.5}$ erg/s. Based on the luminosity after 1 million years, we find that the lifetime ($\tau$) of the $\phi$ bosons must be greater than 1.85 $\times  $10$^{11}$ years. With such a long lifetime the luminosity stays essentially constant.
%
%

In the next section the consequences of $\phi$ boson decay into standard model particles are explored.

\section{Decay modes of $\phi$ bosons}
\label{bosonDecay}
The decay products of the neutron, the $\chi$ and $\phi$ are BSM particles (particles beyond the Standard Model) that can originate from 
some UV complete theory or be considered within some low energy effective theory. While remaining agnostic about their origin we can comment on the constraints on them from a variety of sources. The massive fermion, $\chi$, is an ideal candidate for dark matter and can form the bulk or all of the observed relic density today. We leave a detailed discussion on the details of this mechanism for a later expanded work.
The other product of the decay, the boson, can originate from a BSM source. 
On general grounds and experimental considerations the possibility that the boson is a photon has been ruled out. Here we consider some simple possibilites for the bosons to couple SM particles, and constraints on the basis of findings in the previous sections.

\subsection{Scalars and Pseudoscalars}
In the last few years light scalar and pseudo-scalar particles have emerged as leading new physics candidates that can be constrained 
from a variety of sources. While the primary motivation is derived from 
axions, simplified models with light scalars or pseudo-scalars have triggered a lot of attention. Here we assess their viability 
given our findings above. 

The first bosonic candidate is a scalar coupled to the electromagnetic field strength,   
\begin{equation}
    \mathcal{L}_{int}= \frac{C_{s}}{\Lambda} \phi F_{\mu\nu}F^{\mu\nu} + \frac{m_{f}}{\Lambda} \phi\bar{f}f +\cdots
\end{equation} 
where $\phi$ is the scalar field, $F_{\mu\nu}$  the electromagnetic field strength, and $f$ the Dirac spinor for the leptons. The overall normalization $\frac{C_{s}}{\Lambda}$ is model dependent, while $m_{l}$ is the mass of the lepton. The linear couplings can be generated by from 
the scalar coupling to Higgs, as $\phi H^{\dagger}H$.  
A quadratic coupling can also be generated if $\phi$ carries a $Z_{2}$ symmtery, 
\begin{equation}
\mathcal{L} = \frac{C_{q}}{\Lambda_{q}^{2}} \phi^{2} F_{\mu\nu}F^{\mu\nu} + \sum_{f}\frac{m_{f}}{\Lambda_{q}^{2}}\phi^{2}\bar{f}f +\cdots
\end{equation}
In both cases the dots indicate any other couplings that may be induced.

Couplings to the neutron can be obtained by integrating out, for example, heavy fermions yielding dimension 6 operators, such that the effective neutron coupling can be written as, 
\begin{equation}
    \mathcal{L} \in L_{kin} + \lambda_{eff} n\chi \phi \, .
\end{equation}
The linear (and quadratic) couplings induce a shift in the electromagnetic couplings that can be constrained from a variety of sources. A summary of these can be found in Ref.~\cite{Antypas:2022asj}. 
 
The next possibility is that of a pseudoscalar that couples like an axion (like particle) to photons, and derivatively to electrons 
\begin{equation}
    \mathcal{L}_{int} = \frac{C_{s\gamma}}{\Lambda}\phi F_{\mu\nu}\tilde{F}^{\mu\nu} + \frac{C_{f}}{2\Lambda}(\partial_{\mu}\phi)\bar{f}\gamma^{\mu}\gamma^{5}f  +\cdots
    \label{Eq:alpeff}
\end{equation}
For axion-like particles in Eq.~\ref{Eq:alpeff}, the effective ALP coupling to leptons generates a coupling, 
\begin{equation}
\frac{C_{f}}{2\Lambda}(\partial_{\mu}\phi)\bar{f}\gamma^{\mu}\gamma^{5}f = -\frac{C_{f}m_{f}}{\Lambda} i\bar{f}\gamma^{\mu}\gamma^{5}f +\cdots
\end{equation}
where the dots indicate terms proportional to $F\tilde{F}$. The decay widths to charged fermions are given by, 
\begin{equation}
\Gamma(a\to f\bar{f})=\frac{m_{f}^{2}m_{a}|C_{f}^{2}|}{4\pi \Lambda^{2}}\sqrt{1- \frac{4m_{f}^{2}}{m_{a}^{2}}} \, .
\end{equation}

Analogous to scalars the effective ALP coupling to neutrons can be written as, 
\begin{equation}
    \mathcal{L} \in L_{kin} + \lambda_{eff} n\chi\gamma^{5} 
    \phi \, .
\end{equation}
A comprehensive account of UV complete models and their phenomenological consequences is left for future work.
In principle, since the bosons in our case are heavy, the most general Lagrangian will contain interaction terms involving not only photons and leptons, but hadrons as well. 

The decay widths for pseudoscalars to diphotons are given by,
\begin{equation}
\Gamma(a\to\gamma\gamma) = \frac{|C_{\gamma}^{2}|}{4\pi \Lambda^{2}}m_{a}^{3} \, .
\end{equation}
The lifetime is
\begin{equation}
\tau(a\to\gamma\gamma)= 1/\Gamma(a\to\gamma\gamma\times f) 
\end{equation}
where $f$ is the conversion factor from $\rm GeV^{-1}$ to seconds. From the estimates derived above, for a boson mass of 1 MeV, a lifetime of $\tau\geq\simeq 10^{11}$ years, and if this is only decay channel relevant, 
the effective coupling, $g_{eff}=C_{\gamma}/\Lambda\leq 10^{-17}$. Note that a lifetime of $10^{11}$ years is about $10^{18}$ seconds. The lifetime of the universe is about $10^{18}$ seconds and therefore this boson is cosmologically stable and should add to the total relic density of the universe. The exact amount of dark matter density depends on the co-efficient $C_{f}$, as well as the decay constant $\Lambda$. Typically in axion like models, like the ones considered here, we can obtain a significant fraction of the dark matter with a $\mathcal{O}$(1) misallignement angle. 

There are however significant constraints of models of this class. 
For scalars and pseudoscalars, one of the strongest constraints at this mass originates from the consideration that photons produced during ALP decays when the Universe is transparent should not exceed the total extragalactic background light
(EBL)~\cite{Cadamuro:2011fd}. For pseudoscalar ALPs, this limits lifetimes to $\tau\geq 10^{23}$ seconds,  such that the effective ALP coupling is restricted $g_{eff}\leq 10^{-19}$. Furthermore X-rays produced from ALP decays in galaxies must not exceed the known backgrounds. This limits $\tau\geq 10^{25} $ seconds leading to an effective coupling $g_{eff}\leq 10^{-20}$\cite{Cadamuro:2011fd}.     

\subsection{Spin-1}
While the decay to a photon has been ruled out, a possible solution is that the spin-1 boson can be a dark (kinetically) mixed photon. 
The massless part of the most general theory of two $U(1)_{a,b}$ Abelian gauge bosons can be written as, 
\begin{equation}
\mathcal{L} = -\frac{1}{4} F_{a\mu\nu}F^{\mu\nu}_{a} - \frac{1}{4} F_{b\mu\nu}F^{\mu\nu}_{b} - \frac{\varepsilon}{2} F_{a\mu\nu}F^{\mu\nu}_{b}  
\end{equation}
The masses of these can be obtained via a Stuckelberg mechanism, or via a spontaneously broken gauge symmetry
\begin{equation}
    \mathcal{L}_{m} = \frac{1}{2} M_{a}^{2}A_{\mu}^{a}A^{a\mu} + \frac{1}{2} M_{b}^{2}A_{\mu}^{b}A^{b\mu} + M_{a}M_{b}A_{\mu}^{a} A^{\mu}_{b}
\end{equation}

Consider a hypercharge mixing with the usual photon, 
\begin{equation}
\mathcal{L}= \frac{\epsilon}{2 \cos{\theta}_{W}}\tilde{F}^{'}_{\mu\nu}B^{\mu\nu} \, .
\end{equation}
Then, the effective Lagrangian becomes, 
\begin{equation}
    \mathcal{L}\in e \epsilon J_{\mu}A_{\mu}^{'} + e'\epsilon \tan{\theta}_{W} J_{\mu}^{'}Z_{\mu} + e' J_{\mu}^{'}A_{\mu}^{'} \, ,
\end{equation}
where $J_{\mu}^{'}$ and $e'$ are the dark sector current and the dark photon coupling to the dark sector. Once the $Z$ boson is integrated out we can see that the coupling of the dark photon to SM fermions is proportional to $e\epsilon$, i.e., millicharged dark photons which are constrained from various sources. 
The effective coupling to neutrons can be written as, 
\begin{equation}
    \mathcal{L} \in e\epsilon (n\sigma^{\mu\nu}\chi F^{'}_{\mu\nu} )  \, .  
\end{equation}

Below the two electron threshold, the constraints on dark photons originate from stellar cooling bounds and from the Xenon-1T experiments~\cite{Caputo:2021eaa,PhysRevD.106.022001}. The constraints on the kinetic mixing parameter is 
$\epsilon\leq 10^{-13}$ for $m_{A'}\simeq 1 MeV$.

If the dark photon is extremely light, if produced non-thermally like a condensate, like the axion with a misallignement mechanism.
In this case the mass is generated by the Stuckelberg mechanism, and the generation of the relic follows like the usual axion. 

Additionally, in most relevant models, the dark photon is accompanied to dark fermions. Here dark photon can account for the relic density through a freeze-in mechanism within the dark sector or through feeble couplings to SM. 

\section{Conclusion}
In this work, we explored the consequences of neutron decays into a dark sector inside neutron stars. Working on the hypothesis that $n\to \chi + \phi$, we analyzed the feasibility of this decay by studying the effect of the corresponding equation of state on the properties of the neutron star, including its mass, radius and tidal deformability. We then focused on the possibility that the $\phi$ remains trapped inside the star leading to heating. We concluded that if the $\phi$ has a mass of around 1 MeV, based upon observations of luminosity of stars as a function of age, the $\phi$ must have a lifetime greater than $10^{11}$ years. Finally we studied the consequences and of a $\phi$ coupled to standard model particles within simplified ALP like models. An expanded work with cosmological consequences, as well as a study of UV completions, is left for future work.

\label{conclusion}

\acknowledgments

This study has been supported by a University of Adelaide International Scholarship (WH) and by the Australian Research Council through the ARC Centre for Dark Matter Particle Physics (CE200100008).

\bibliography{main.bib}

\begin{thebibliography}{999}

\bibitem[Wietfeldt and Greene(2011)]{RevModPhys.83.1173}
Wietfeldt, F.E.; Greene, G.L.
\newblock Colloquium: The neutron lifetime.
\newblock {\em Rev. Mod. Phys.} {\bf 2011}, {\em 83},~1173--1192.
\newblock {\url{https://doi.org/10.1103/RevModPhys.83.1173}}.

\bibitem[Serebrov \em{et~al.}(2018)Serebrov et~al.]{Serebrov:2017bzo}
Serebrov, A.P.;  et~al.
\newblock {Neutron lifetime measurements with a large gravitational trap for
  ultracold neutrons}.
\newblock {\em Phys. Rev. C} {\bf 2018}, {\em 97},~055503,
  \href{http://xxx.lanl.gov/abs/1712.05663}{{\normalfont
  [arXiv:nucl-ex/1712.05663]}}.
\newblock {\url{https://doi.org/10.1103/PhysRevC.97.055503}}.

\bibitem[Pattie \em{et~al.}(2018)Pattie et~al.]{Pattie:2017vsj}
Pattie, Jr., R.W.;  et~al.
\newblock {Measurement of the neutron lifetime using a magneto-gravitational
  trap and in situ detection}.
\newblock {\em Science} {\bf 2018}, {\em 360},~627--632,
  \href{http://xxx.lanl.gov/abs/1707.01817}{{\normalfont
  [arXiv:nucl-ex/1707.01817]}}.
\newblock {\url{https://doi.org/10.1126/science.aan8895}}.

\bibitem[Tan(2019)]{TAN2019134921}
Tan, W.
\newblock Neutron oscillations for solving neutron lifetime and dark matter
  puzzles.
\newblock {\em Physics Letters B} {\bf 2019}, {\em 797},~134921.
\newblock
  {\url{https://doi.org/https://doi.org/10.1016/j.physletb.2019.134921}}.

\bibitem[Steyerl \em{et~al.}(2012)Steyerl, Pendlebury, Kaufman, Malik, and
  Desai]{PhysRevC.85.065503}
Steyerl, A.; Pendlebury, J.M.; Kaufman, C.; Malik, S.S.; Desai, A.M.
\newblock Quasielastic scattering in the interaction of ultracold neutrons with
  a liquid wall and application in a reanalysis of the Mambo I neutron-lifetime
  experiment.
\newblock {\em Phys. Rev. C} {\bf 2012}, {\em 85},~065503.
\newblock {\url{https://doi.org/10.1103/PhysRevC.85.065503}}.

\bibitem[Yue \em{et~al.}(2013)Yue, Dewey, Gilliam, Greene, Laptev, Nico, Snow,
  and Wietfeldt]{PhysRevLett.111.222501}
Yue, A.T.; Dewey, M.S.; Gilliam, D.M.; Greene, G.L.; Laptev, A.B.; Nico, J.S.;
  Snow, W.M.; Wietfeldt, F.E.
\newblock Improved Determination of the Neutron Lifetime.
\newblock {\em Phys. Rev. Lett.} {\bf 2013}, {\em 111},~222501.
\newblock {\url{https://doi.org/10.1103/PhysRevLett.111.222501}}.

\bibitem[Otono(2017)]{Otono:2016fsv}
Otono, H.
\newblock {LiNA \textendash{} Lifetime of neutron apparatus with time
  projection chamber and solenoid coil}.
\newblock {\em Nucl. Instrum. Meth. A} {\bf 2017}, {\em 845},~278--280,
  \href{http://xxx.lanl.gov/abs/1603.06572}{{\normalfont
  [arXiv:physics.ins-det/1603.06572]}}.
\newblock {\url{https://doi.org/10.1016/j.nima.2016.05.042}}.

\bibitem[Olive(2016)]{Olive_2016}
Olive, K.
\newblock Review of Particle Physics.
\newblock {\em Chinese Physics C} {\bf 2016}, {\em 40},~100001.
\newblock {\url{https://doi.org/10.1088/1674-1137/40/10/100001}}.

\bibitem[Gonzalez \em{et~al.}(2021)Gonzalez et~al.]{UCNt:2021pcg}
Gonzalez, F.M.;  et~al.
\newblock {Improved Neutron Lifetime Measurement with UCN\ensuremath{\tau}}.
\newblock {\em Phys. Rev. Lett.} {\bf 2021}, {\em 127},~162501,
  \href{http://xxx.lanl.gov/abs/2106.10375}{{\normalfont
  [arXiv:nucl-ex/2106.10375]}}.
\newblock {\url{https://doi.org/10.1103/PhysRevLett.127.162501}}.

\bibitem[Pattie \em{et~al.}(2018)Pattie, Callahan, Cude-Woods, Adamek,
  Broussard, Clayton, Currie, Dees, Ding, Engel, Fellers, Fox, Geltenbort,
  Hickerson, Hoffbauer, Holley, Komives, Liu, MacDonald, Makela, Morris, Ortiz,
  Ramsey, Salvat, Saunders, Seestrom, Sharapov, Sjue, Tang, Vanderwerp,
  Vogelaar, Walstrom, Wang, Wei, Weaver, Wexler, Womack, Young, and
  Zeck]{doi:10.1126/science.aan8895}
Pattie, R.W.; Callahan, N.B.; Cude-Woods, C.; Adamek, E.R.; Broussard, L.J.;
  Clayton, S.M.; Currie, S.A.; Dees, E.B.; Ding, X.; Engel, E.M.;  et~al.
\newblock Measurement of the neutron lifetime using a magneto-gravitational
  trap and in situ detection.
\newblock {\em Science} {\bf 2018}, {\em 360},~627--632,
  \href{http://xxx.lanl.gov/abs/https://www.science.org/doi/pdf/10.1126/science.aan8895}{{\normalfont
  [https://www.science.org/doi/pdf/10.1126/science.aan8895]}}.
\newblock {\url{https://doi.org/10.1126/science.aan8895}}.

\bibitem[Serebrov \em{et~al.}(2018)Serebrov, Kolomensky, Fomin,
  Krasnoshchekova, Vassiljev, Prudnikov, Shoka, Chechkin, Chaikovskiy,
  Varlamov, Ivanov, Pirozhkov, Geltenbort, Zimmer, Jenke, Van~der Grinten, and
  Tucker]{PhysRevC.97.055503}
Serebrov, A.P.; Kolomensky, E.A.; Fomin, A.K.; Krasnoshchekova, I.A.;
  Vassiljev, A.V.; Prudnikov, D.M.; Shoka, I.V.; Chechkin, A.V.; Chaikovskiy,
  M.E.; Varlamov, V.E.;  et~al.
\newblock Neutron lifetime measurements with a large gravitational trap for
  ultracold neutrons.
\newblock {\em Phys. Rev. C} {\bf 2018}, {\em 97},~055503.
\newblock {\url{https://doi.org/10.1103/PhysRevC.97.055503}}.

\bibitem[Fornal and Grinstein(2018)]{Fornal:2018eol}
Fornal, B.; Grinstein, B.
\newblock {Dark Matter Interpretation of the Neutron Decay Anomaly}.
\newblock {\em Phys. Rev. Lett.} {\bf 2018}, {\em 120},~191801,
  \href{http://xxx.lanl.gov/abs/1801.01124}{{\normalfont
  [arXiv:hep-ph/1801.01124]}}.
\newblock [Erratum: Phys.Rev.Lett. 124, 219901 (2020)],
  {\url{https://doi.org/10.1103/PhysRevLett.120.191801}}.

\bibitem[Grinstein \em{et~al.}(2019)Grinstein, Kouvaris, and
  Nielsen]{Grinstein:2018ptl}
Grinstein, B.; Kouvaris, C.; Nielsen, N.G.
\newblock {Neutron Star Stability in Light of the Neutron Decay Anomaly}.
\newblock {\em Phys. Rev. Lett.} {\bf 2019}, {\em 123},~091601,
  \href{http://xxx.lanl.gov/abs/1811.06546}{{\normalfont
  [arXiv:hep-ph/1811.06546]}}.
\newblock {\url{https://doi.org/10.1103/PhysRevLett.123.091601}}.

\bibitem[Fornal and Grinstein(2020)]{Fornal:2020gto}
Fornal, B.; Grinstein, B.
\newblock {Neutron\textquoteright{}s dark secret}.
\newblock {\em Mod. Phys. Lett. A} {\bf 2020}, {\em 35},~2030019,
  \href{http://xxx.lanl.gov/abs/2007.13931}{{\normalfont
  [arXiv:hep-ph/2007.13931]}}.
\newblock {\url{https://doi.org/10.1142/S0217732320300190}}.

\bibitem[Tang \em{et~al.}(2018)Tang et~al.]{Tang:2018eln}
Tang, Z.;  et~al.
\newblock {Search for the Neutron Decay n$\rightarrow$ X+$\gamma$ where X is a
  dark matter particle}.
\newblock {\em Phys. Rev. Lett.} {\bf 2018}, {\em 121},~022505,
  \href{http://xxx.lanl.gov/abs/1802.01595}{{\normalfont
  [arXiv:nucl-ex/1802.01595]}}.
\newblock {\url{https://doi.org/10.1103/PhysRevLett.121.022505}}.

\bibitem[Serebrov \em{et~al.}(2008)Serebrov et~al.]{Serebrov:2007gw}
Serebrov, A.P.;  et~al.
\newblock {Experimental search for neutron: Mirror neutron oscillations using
  storage of ultracold neutrons}.
\newblock {\em Phys. Lett. B} {\bf 2008}, {\em 663},~181--185,
  \href{http://xxx.lanl.gov/abs/0706.3600}{{\normalfont
  [arXiv:nucl-ex/0706.3600]}}.
\newblock {\url{https://doi.org/10.1016/j.physletb.2008.04.014}}.

\bibitem[Motta \em{et~al.}(2018{\natexlab{a}})Motta, Guichon, and
  Thomas]{Motta:2018rxp}
Motta, T.F.; Guichon, P.A.M.; Thomas, A.W.
\newblock {Implications of Neutron Star Properties for the Existence of Light
  Dark Matter}.
\newblock {\em J. Phys. G} {\bf 2018}, {\em 45},~05LT01,
  \href{http://xxx.lanl.gov/abs/1802.08427}{{\normalfont
  [arXiv:nucl-th/1802.08427]}}.
\newblock {\url{https://doi.org/10.1088/1361-6471/aab689}}.

\bibitem[Motta \em{et~al.}(2018{\natexlab{b}})Motta, Guichon, and
  Thomas]{Motta:2018bil}
Motta, T.F.; Guichon, P.A.M.; Thomas, A.W.
\newblock {Neutron to Dark Matter Decay in Neutron Stars}.
\newblock {\em Int. J. Mod. Phys. A} {\bf 2018}, {\em 33},~1844020,
  \href{http://xxx.lanl.gov/abs/1806.00903}{{\normalfont
  [arXiv:nucl-th/1806.00903]}}.
\newblock {\url{https://doi.org/10.1142/S0217751X18440207}}.

\bibitem[Baym \em{et~al.}(2018)Baym, Beck, Geltenbort, and
  Shelton]{PhysRevLett.121.061801}
Baym, G.; Beck, D.H.; Geltenbort, P.; Shelton, J.
\newblock Testing Dark Decays of Baryons in Neutron Stars.
\newblock {\em Phys. Rev. Lett.} {\bf 2018}, {\em 121},~061801.
\newblock {\url{https://doi.org/10.1103/PhysRevLett.121.061801}}.

\bibitem[McKeen \em{et~al.}(2018)McKeen, Nelson, Reddy, and Zhou]{2018_sa}
McKeen, D.; Nelson, A.E.; Reddy, S.; Zhou, D.
\newblock Neutron Stars Exclude Light Dark Baryons.
\newblock {\em Physical Review Letters} {\bf 2018}, {\em 121}.
\newblock {\url{https://doi.org/10.1103/physrevlett.121.061802}}.

\bibitem[Husain \em{et~al.}(2022)Husain, Motta, and Thomas]{Husain:2022bxl}
Husain, W.; Motta, T.F.; Thomas, A.W.
\newblock {Consequences of neutron decay inside neutron stars}.
\newblock {\em JCAP} {\bf 2022}, {\em 10},~028,
  \href{http://xxx.lanl.gov/abs/2203.02758}{{\normalfont
  [arXiv:hep-ph/2203.02758]}}.
\newblock {\url{https://doi.org/10.1088/1475-7516/2022/10/028}}.

\bibitem[Ivanov \em{et~al.}(2018)Ivanov, H\"ollwieser, Troitskaya, Wellenzohn,
  and Berdnikov]{Ivanov:2018vit}
Ivanov, A.N.; H\"ollwieser, R.; Troitskaya, N.I.; Wellenzohn, M.; Berdnikov,
  Y.A.
\newblock {Neutron Dark Matter Decays} {\bf 2018}.
\newblock  \href{http://xxx.lanl.gov/abs/1806.10107}{{\normalfont
  [arXiv:hep-ph/1806.10107]}}.

\bibitem[Strumia(2021)]{Strumia:2021ybk}
Strumia, A.
\newblock {Dark Matter interpretation of the neutron decay anomaly} {\bf 2021}.
\newblock  \href{http://xxx.lanl.gov/abs/2112.09111}{{\normalfont
  [arXiv:hep-ph/2112.09111]}}.

\bibitem[Husain and Thomas(2022)]{Husain:2022brl}
Husain, W.; Thomas, A.W.
\newblock {Novel neutron decay mode inside neutron stars} {\bf 2022}.
\newblock  \href{http://xxx.lanl.gov/abs/2206.11262}{{\normalfont
  [arXiv:hep-ph/2206.11262]}}.

\bibitem[Mukhopadhyay \em{et~al.}(2017)Mukhopadhyay, Atta, Imam, Basu, and
  Samanta]{Mukhopadhyay_2017}
Mukhopadhyay, S.; Atta, D.; Imam, K.; Basu, D.N.; Samanta, C.
\newblock Compact bifluid hybrid stars: hadronic matter mixed with
  self-interacting fermionic asymmetric dark matter.
\newblock {\em The European Physical Journal C} {\bf 2017}, {\em 77}.
\newblock {\url{https://doi.org/10.1140/epjc/s10052-017-5006-3}}.

\bibitem[Bertone and Fairbairn(2008)]{PhysRevD.77.043515}
Bertone, G.; Fairbairn, M.
\newblock Compact stars as dark matter probes.
\newblock {\em Phys. Rev. D} {\bf 2008}, {\em 77},~043515.
\newblock {\url{https://doi.org/10.1103/PhysRevD.77.043515}}.

\bibitem[Kouvaris(2008)]{PhysRevD.77.023006}
Kouvaris, C.
\newblock WIMP annihilation and cooling of neutron stars.
\newblock {\em Phys. Rev. D} {\bf 2008}, {\em 77},~023006.
\newblock {\url{https://doi.org/10.1103/PhysRevD.77.023006}}.

\bibitem[Ciarcelluti and Sandin(2011)]{CIARCELLUTI201119}
Ciarcelluti, P.; Sandin, F.
\newblock Have neutron stars a dark matter core?
\newblock {\em Physics Letters B} {\bf 2011}, {\em 695},~19--21.
\newblock
  {\url{https://doi.org/https://doi.org/10.1016/j.physletb.2010.11.021}}.

\bibitem[Sandin and Ciarcelluti(2009)]{Sandin_2009}
Sandin, F.; Ciarcelluti, P.
\newblock Effects of mirror dark matter on neutron stars.
\newblock {\em Astroparticle Physics} {\bf 2009}, {\em 32},~278–284.
\newblock {\url{https://doi.org/10.1016/j.astropartphys.2009.09.005}}.

\bibitem[Leung \em{et~al.}(2011)Leung, Chu, and Lin]{Leung:2011zz}
Leung, S.C.; Chu, M.C.; Lin, L.M.
\newblock {Dark-matter admixed neutron stars}.
\newblock {\em Phys. Rev. D} {\bf 2011}, {\em 84},~107301,
  \href{http://xxx.lanl.gov/abs/1111.1787}{{\normalfont
  [arXiv:astro-ph.CO/1111.1787]}}.
\newblock {\url{https://doi.org/10.1103/PhysRevD.84.107301}}.

\bibitem[Ellis \em{et~al.}(2018)Ellis, Hütsi, Kannike, Marzola, Raidal, and
  Vaskonen]{Ellis_2018}
Ellis, J.; Hütsi, G.; Kannike, K.; Marzola, L.; Raidal, M.; Vaskonen, V.
\newblock Dark matter effects on neutron star properties.
\newblock {\em Physical Review D} {\bf 2018}, {\em 97}.
\newblock {\url{https://doi.org/10.1103/physrevd.97.123007}}.

\bibitem[Bell \em{et~al.}(2020)Bell, Busoni, Motta, Robles, Thomas, and
  Virgato]{bell2020nucleon}
Bell, N.F.; Busoni, G.; Motta, T.F.; Robles, S.; Thomas, A.W.; Virgato, M.
\newblock Nucleon Structure and Strong Interactions in Dark Matter Capture in
  Neutron Stars,  2020,  \href{http://xxx.lanl.gov/abs/2012.08918}{{\normalfont
  [arXiv:hep-ph/2012.08918]}}.

\bibitem[Husain and Thomas(2021)]{2021_w}
Husain, W.; Thomas, A.W.
\newblock Possible nature of dark matter.
\newblock {\em Journal of Cosmology and Astroparticle Physics} {\bf 2021}, {\em
  2021},~086.
\newblock {\url{https://doi.org/10.1088/1475-7516/2021/10/086}}.

\bibitem[{Mielke} and {Schunck}(2000)]{2000NuPhB.564..185M}
{Mielke}, E.W.; {Schunck}, F.E.
\newblock {Boson stars: alternatives to primordial black holes?}
\newblock {\em Nuclear Physics B} {\bf 2000}, {\em 1},~185--203,
  \href{http://xxx.lanl.gov/abs/gr-qc/0001061}{{\normalfont
  [arXiv:gr-qc/gr-qc/0001061]}}.
\newblock {\url{https://doi.org/10.1016/S0550-3213(99)00492-7}}.

\bibitem[Blinnikov and Khlopov(1983)]{Blinnikov:1983gh}
Blinnikov, S.I.; Khlopov, M.
\newblock {Possible astronomical effects of mirror particles}.
\newblock {\em Sov. Astron.} {\bf 1983}, {\em 27},~371--375.

\bibitem[Horowitz and Reddy(2019)]{2019_reddy}
Horowitz, C.; Reddy, S.
\newblock Gravitational Waves from Compact Dark Objects in Neutron Stars.
\newblock {\em Physical Review Letters} {\bf 2019}, {\em 122}.
\newblock {\url{https://doi.org/10.1103/physrevlett.122.071102}}.

\bibitem[Bertoni \em{et~al.}(2013)Bertoni, Nelson, and Reddy]{2013_red}
Bertoni, B.; Nelson, A.E.; Reddy, S.
\newblock Dark matter thermalization in neutron stars.
\newblock {\em Physical Review D} {\bf 2013}, {\em 88}.
\newblock {\url{https://doi.org/10.1103/physrevd.88.123505}}.

\bibitem[Berryman \em{et~al.}(2022)Berryman, Gardner, and
  Zakeri]{berryman2022neutron}
Berryman, J.M.; Gardner, S.; Zakeri, M.
\newblock Neutron Stars with Baryon Number Violation, Probing Dark Sectors,
  2022,  \href{http://xxx.lanl.gov/abs/2201.02637}{{\normalfont
  [arXiv:hep-ph/2201.02637]}}.

\bibitem[McKeen \em{et~al.}(2021)McKeen, Pospelov, and Raj]{McKeen:2021jbh}
McKeen, D.; Pospelov, M.; Raj, N.
\newblock {Neutron Star Internal Heating Constraints on Mirror Matter}.
\newblock {\em Phys. Rev. Lett.} {\bf 2021}, {\em 127},~061805,
  \href{http://xxx.lanl.gov/abs/2105.09951}{{\normalfont
  [arXiv:hep-ph/2105.09951]}}.
\newblock {\url{https://doi.org/10.1103/PhysRevLett.127.061805}}.

\bibitem[de~Lavallaz and Fairbairn(2010)]{deLavallaz:2010wp}
de~Lavallaz, A.; Fairbairn, M.
\newblock {Neutron Stars as Dark Matter Probes}.
\newblock {\em Phys. Rev. D} {\bf 2010}, {\em 81},~123521,
  \href{http://xxx.lanl.gov/abs/1004.0629}{{\normalfont
  [arXiv:astro-ph.GA/1004.0629]}}.
\newblock {\url{https://doi.org/10.1103/PhysRevD.81.123521}}.

\bibitem[Busoni(2021)]{Busoni:2021zoe}
Busoni, G.
\newblock {Capture of Dark Matter in Neutron Stars} {\bf 2021}.
\newblock  \href{http://xxx.lanl.gov/abs/2201.00048}{{\normalfont
  [arXiv:hep-ph/2201.00048]}}.

\bibitem[Sen and Guha(2021)]{Sen:2021wev}
Sen, D.; Guha, A.
\newblock {Implications of feebly interacting dark sector on neutron star
  properties and constraints from GW170817}.
\newblock {\em Mon. Not. Roy. Astron. Soc.} {\bf 2021}, {\em 504},~3,
  \href{http://xxx.lanl.gov/abs/2104.06141}{{\normalfont
  [arXiv:hep-ph/2104.06141]}}.
\newblock {\url{https://doi.org/10.1093/mnras/stab1056}}.

\bibitem[Guha and Sen(2021)]{Guha:2021njn}
Guha, A.; Sen, D.
\newblock {Feeble DM-SM interaction via new scalar and vector mediators in
  rotating neutron stars}.
\newblock {\em JCAP} {\bf 2021}, {\em 09},~027,
  \href{http://xxx.lanl.gov/abs/2106.10353}{{\normalfont
  [arXiv:hep-ph/2106.10353]}}.
\newblock {\url{https://doi.org/10.1088/1475-7516/2021/09/027}}.

\bibitem[Abbott \em{et~al.}(2018)Abbott et~al.]{LIGOScientific:2018cki}
Abbott, B.P.;  et~al.
\newblock {GW170817: Measurements of neutron star radii and equation of state}.
\newblock {\em Phys. Rev. Lett.} {\bf 2018}, {\em 121},~161101,
  \href{http://xxx.lanl.gov/abs/1805.11581}{{\normalfont
  [arXiv:gr-qc/1805.11581]}}.
\newblock {\url{https://doi.org/10.1103/PhysRevLett.121.161101}}.

\bibitem[{Demorest} \em{et~al.}(2010){Demorest}, {Pennucci}, {Ransom},
  {Roberts}, and {Hessels}]{2010Natur.467.1081D}
{Demorest}, P.B.; {Pennucci}, T.; {Ransom}, S.M.; {Roberts}, M.S.E.; {Hessels},
  J.W.T.
\newblock {A two-solar-mass neutron star measured using Shapiro delay}.
\newblock {\em nat} {\bf 2010}, {\em 467},~1081--1083,
  \href{http://xxx.lanl.gov/abs/1010.5788}{{\normalfont
  [arXiv:astro-ph.HE/1010.5788]}}.
\newblock {\url{https://doi.org/10.1038/nature09466}}.

\bibitem[{Antoniadis} \em{et~al.}(2013){Antoniadis}, {Freire}, {Wex}, {Tauris},
  {Lynch}, {van Kerkwijk}, {Kramer}, {Bassa}, {Dhillon}, {Driebe}, {Hessels},
  {Kaspi}, {Kondratiev}, {Langer}, {Marsh}, {McLaughlin}, {Pennucci}, {Ransom},
  {Stairs}, {van Leeuwen}, {Verbiest}, and {Whelan}]{2013Sci...340..448A}
{Antoniadis}, J.; {Freire}, P.C.C.; {Wex}, N.; {Tauris}, T.M.; {Lynch}, R.S.;
  {van Kerkwijk}, M.H.; {Kramer}, M.; {Bassa}, C.; {Dhillon}, V.S.; {Driebe},
  T.;  et~al.
\newblock {A Massive Pulsar in a Compact Relativistic Binary}.
\newblock {\em Science} {\bf 2013}, {\em 340},~448,
  \href{http://xxx.lanl.gov/abs/1304.6875}{{\normalfont
  [arXiv:astro-ph.HE/1304.6875]}}.
\newblock {\url{https://doi.org/10.1126/science.1233232}}.

\bibitem[Riley \em{et~al.}(2019)Riley et~al.]{Riley:2019yda}
Riley, T.E.;  et~al.
\newblock {A $NICER$ View of PSR J0030+0451: Millisecond Pulsar Parameter
  Estimation}.
\newblock {\em Astrophys. J. Lett.} {\bf 2019}, {\em 887},~L21,
  \href{http://xxx.lanl.gov/abs/1912.05702}{{\normalfont
  [arXiv:astro-ph.HE/1912.05702]}}.
\newblock {\url{https://doi.org/10.3847/2041-8213/ab481c}}.

\bibitem[Riley \em{et~al.}(2021)Riley et~al.]{Riley:2021pdl}
Riley, T.E.;  et~al.
\newblock {A NICER View of the Massive Pulsar PSR J0740+6620 Informed by Radio
  Timing and XMM-Newton Spectroscopy}.
\newblock {\em Astrophys. J. Lett.} {\bf 2021}, {\em 918},~L27,
  \href{http://xxx.lanl.gov/abs/2105.06980}{{\normalfont
  [arXiv:astro-ph.HE/2105.06980]}}.
\newblock {\url{https://doi.org/10.3847/2041-8213/ac0a81}}.

\bibitem[Miller \em{et~al.}(2019)Miller et~al.]{Miller:2019cac}
Miller, M.C.;  et~al.
\newblock {PSR J0030+0451 Mass and Radius from $NICER$ Data and Implications
  for the Properties of Neutron Star Matter}.
\newblock {\em Astrophys. J. Lett.} {\bf 2019}, {\em 887},~L24,
  \href{http://xxx.lanl.gov/abs/1912.05705}{{\normalfont
  [arXiv:astro-ph.HE/1912.05705]}}.
\newblock {\url{https://doi.org/10.3847/2041-8213/ab50c5}}.

\bibitem[Miller \em{et~al.}(2021)Miller et~al.]{Miller:2021qha}
Miller, M.C.;  et~al.
\newblock {The Radius of PSR J0740+6620 from NICER and XMM-Newton Data}.
\newblock {\em Astrophys. J. Lett.} {\bf 2021}, {\em 918},~L28,
  \href{http://xxx.lanl.gov/abs/2105.06979}{{\normalfont
  [arXiv:astro-ph.HE/2105.06979]}}.
\newblock {\url{https://doi.org/10.3847/2041-8213/ac089b}}.

\bibitem[Abbott \em{et~al.}(2017)Abbott, Abbott, Abbott, Acernese, Ackley,
  Adams, Adams, Addesso, Adhikari, Adya, and et~al.]{Abbott_2017}
Abbott, B.; Abbott, R.; Abbott, T.; Acernese, F.; Ackley, K.; Adams, C.; Adams,
  T.; Addesso, P.; Adhikari, R.; Adya, V.;  et~al.
\newblock GW170817: Observation of Gravitational Waves from a Binary Neutron
  Star Inspiral.
\newblock {\em Physical Review Letters} {\bf 2017}, {\em 119}.
\newblock {\url{https://doi.org/10.1103/physrevlett.119.161101}}.

\bibitem[Abbott \em{et~al.}(2019)Abbott, Abbott, Abbott, Abraham, Acernese,
  Ackley, Adams, Adhikari, Adya, Affeldt, and et~al.]{Abbott_2019}
Abbott, B.; Abbott, R.; Abbott, T.; Abraham, S.; Acernese, F.; Ackley, K.;
  Adams, C.; Adhikari, R.; Adya, V.; Affeldt, C.;  et~al.
\newblock GWTC-1: A Gravitational-Wave Transient Catalog of Compact Binary
  Mergers Observed by LIGO and Virgo during the First and Second Observing
  Runs.
\newblock {\em Physical Review X} {\bf 2019}, {\em 9}.
\newblock {\url{https://doi.org/10.1103/physrevx.9.031040}}.

\bibitem[Lawley \em{et~al.}(2006)Lawley, Bentz, and Thomas]{Lawley:2006ps}
Lawley, S.; Bentz, W.; Thomas, A.W.
\newblock {Nucleons, nuclear matter and quark matter: A Unified NJL approach}.
\newblock {\em J. Phys. G} {\bf 2006}, {\em 32},~667--680,
  \href{http://xxx.lanl.gov/abs/nucl-th/0602014}{{\normalfont
  [nucl-th/0602014]}}.
\newblock {\url{https://doi.org/10.1088/0954-3899/32/5/006}}.

\bibitem[Whittenbury \em{et~al.}(2014)Whittenbury, Carroll, Thomas, Tsushima,
  and Stone]{Whittenbury:2013wma}
Whittenbury, D.L.; Carroll, J.D.; Thomas, A.W.; Tsushima, K.; Stone, J.R.
\newblock {Quark-Meson Coupling Model, Nuclear Matter Constraints and Neutron
  Star Properties}.
\newblock {\em Phys. Rev. C} {\bf 2014}, {\em 89},~065801,
  \href{http://xxx.lanl.gov/abs/1307.4166}{{\normalfont
  [arXiv:nucl-th/1307.4166]}}.
\newblock {\url{https://doi.org/10.1103/PhysRevC.89.065801}}.

\bibitem[Whittenbury \em{et~al.}(2016)Whittenbury, Matevosyan, and
  Thomas]{Whittenbury:2015ziz}
Whittenbury, D.L.; Matevosyan, H.H.; Thomas, A.W.
\newblock {Hybrid stars using the quark-meson coupling and proper-time
  Nambu\textendash{}Jona-Lasinio models}.
\newblock {\em Phys. Rev. C} {\bf 2016}, {\em 93},~035807,
  \href{http://xxx.lanl.gov/abs/1511.08561}{{\normalfont
  [arXiv:nucl-th/1511.08561]}}.
\newblock {\url{https://doi.org/10.1103/PhysRevC.93.035807}}.

\bibitem[Bodmer(1971)]{PhysRevD.4.1601}
Bodmer, A.R.
\newblock Collapsed Nuclei.
\newblock {\em Phys. Rev. D} {\bf 1971}, {\em 4},~1601--1606.
\newblock {\url{https://doi.org/10.1103/PhysRevD.4.1601}}.

\bibitem[Witten(1984)]{PhysRevD.30.272}
Witten, E.
\newblock Cosmic separation of phases.
\newblock {\em Phys. Rev. D} {\bf 1984}, {\em 30},~272--285.
\newblock {\url{https://doi.org/10.1103/PhysRevD.30.272}}.

\bibitem[Bombaci \em{et~al.}(2004)Bombaci, Parenti, and Vidana]{Bombaci_2004}
Bombaci, I.; Parenti, I.; Vidana, I.
\newblock Quark Deconfinement and Implications for the Radius and the Limiting
  Mass of Compact Stars.
\newblock {\em The Astrophysical Journal} {\bf 2004}, {\em 614},~314--325.
\newblock {\url{https://doi.org/10.1086/423658}}.

\bibitem[Ren and Zhang(2020)]{PhysRevD.102.083003}
Ren, J.; Zhang, C.
\newblock Quantum nucleation of up-down quark matter and astrophysical
  implications.
\newblock {\em Phys. Rev. D} {\bf 2020}, {\em 102},~083003.
\newblock {\url{https://doi.org/10.1103/PhysRevD.102.083003}}.

\bibitem[Terazawa(1989)]{doi:10.1143/JPSJ.58.3555}
Terazawa, H.
\newblock Super-Hypernuclei in the Quark-Shell Model.
\newblock {\em Journal of the Physical Society of Japan} {\bf 1989}, {\em
  58},~3555--3563,
  \href{http://xxx.lanl.gov/abs/https://doi.org/10.1143/JPSJ.58.3555}{{\normalfont
  [https://doi.org/10.1143/JPSJ.58.3555]}}.
\newblock {\url{https://doi.org/10.1143/JPSJ.58.3555}}.

\bibitem[Bednarek \em{et~al.}(2012)Bednarek, Haensel, Zdunik, Bejger, and
  Mańka]{2012_a}
Bednarek, I.; Haensel, P.; Zdunik, J.L.; Bejger, M.; Mańka, R.
\newblock Hyperons in neutron-star cores and a 2M pulsar.
\newblock {\em Astronomy \& Astrophysics} {\bf 2012}, {\em 543},~A157.
\newblock {\url{https://doi.org/10.1051/0004-6361/201118560}}.

\bibitem[Vidaña(2015)]{doi:10.1063/1.4909561}
Vidaña, I.
\newblock Hyperons and neutron stars.
\newblock {\em AIP Conference Proceedings} {\bf 2015}, {\em 1645},~79--85,
  \href{http://xxx.lanl.gov/abs/https://aip.scitation.org/doi/pdf/10.1063/1.4909561}{{\normalfont
  [https://aip.scitation.org/doi/pdf/10.1063/1.4909561]}}.
\newblock {\url{https://doi.org/10.1063/1.4909561}}.

\bibitem[Oertel \em{et~al.}(2016)Oertel, Gulminelli, Providência, and
  Raduta]{2016_a}
Oertel, M.; Gulminelli, F.; Providência, C.; Raduta, A.R.
\newblock Hyperons in neutron stars and supernova cores.
\newblock {\em The European Physical Journal A} {\bf 2016}, {\em 52}.
\newblock {\url{https://doi.org/10.1140/epja/i2016-16050-1}}.

\bibitem[Akmal \em{et~al.}(1998)Akmal, Pandharipande, and
  Ravenhall]{PhysRevC.58.1804}
Akmal, A.; Pandharipande, V.R.; Ravenhall, D.G.
\newblock Equation of state of nucleon matter and neutron star structure.
\newblock {\em Phys. Rev. C} {\bf 1998}, {\em 58},~1804--1828.
\newblock {\url{https://doi.org/10.1103/PhysRevC.58.1804}}.

\bibitem[Balberg and Gal(1997)]{BALBERG1997435}
Balberg, S.; Gal, A.
\newblock An effective equation of state for dense matter with strangeness.
\newblock {\em Nuclear Physics A} {\bf 1997}, {\em 625},~435--472.
\newblock
  {\url{https://doi.org/https://doi.org/10.1016/S0375-9474(97)81465-0}}.

\bibitem[{Glendenning}(1985)]{1985ApJ...293..470G}
{Glendenning}, N.K.
\newblock {Neutron stars are giant hypernuclei ?}
\newblock {\em apj} {\bf 1985}, {\em 293},~470--493.
\newblock {\url{https://doi.org/10.1086/163253}}.

\bibitem[Kaplan and Nelson(1986)]{KAPLAN198657}
Kaplan, D.; Nelson, A.
\newblock Strange goings on in dense nucleonic matter.
\newblock {\em Physics Letters B} {\bf 1986}, {\em 175},~57--63.
\newblock {\url{https://doi.org/https://doi.org/10.1016/0370-2693(86)90331-X}}.

\bibitem[Glendenning \em{et~al.}(1997)Glendenning, Pei, and
  Weber]{PhysRevLett.79.1603}
Glendenning, N.K.; Pei, S.; Weber, F.
\newblock Signal of Quark Deconfinement in the Timing Structure of Pulsar
  Spin-Down.
\newblock {\em Phys. Rev. Lett.} {\bf 1997}, {\em 79},~1603--1606.
\newblock {\url{https://doi.org/10.1103/PhysRevLett.79.1603}}.

\bibitem[Glendenning and Moszkowski(1991)]{PhysRevLett.67.2414}
Glendenning, N.K.; Moszkowski, S.A.
\newblock Reconciliation of neutron-star masses and binding of the
  \ensuremath{\Lambda} in hypernuclei.
\newblock {\em Phys. Rev. Lett.} {\bf 1991}, {\em 67},~2414--2417.
\newblock {\url{https://doi.org/10.1103/PhysRevLett.67.2414}}.

\bibitem[Glendenning(1997)]{Glendenning1997}
Glendenning, N.K., Quark Stars.
\newblock In {\em Compact Stars: Nuclear Physics, Particle Physics and General
  Relativity}; Springer US: New York, NY,  1997; pp. 289--302.
\newblock {\url{https://doi.org/10.1007/978-1-4684-0491-3_8}}.

\bibitem[Haensel and Zdunik(2017)]{Haensel2017}
Haensel, P.; Zdunik, J.L., Nuclear Matter in Neutron Stars.
\newblock In {\em Handbook of Supernovae}; Alsabti, A.W.; Murdin, P., Eds.;
  Springer International Publishing: Cham,  2017; pp. 1331--1351.
\newblock {\url{https://doi.org/10.1007/978-3-319-21846-5_68}}.

\bibitem[{Shuryak}(1980)]{1980PhR....61...71S}
{Shuryak}, E.V.
\newblock {Quantum chromodynamics and the theory of superdense matter}.
\newblock {\em physrep} {\bf 1980}, {\em 61},~71--158.
\newblock {\url{https://doi.org/10.1016/0370-1573(80)90105-2}}.

\bibitem[Weber \em{et~al.}(2007)Weber, Negreiros, and
  Rosenfield]{weber2007neutron}
Weber, F.; Negreiros, R.; Rosenfield, P.
\newblock Neutron Star Interiors and the Equation of State of Superdense
  Matter,  2007,  \href{http://xxx.lanl.gov/abs/0705.2708}{{\normalfont
  [arXiv:astro-ph/0705.2708]}}.

\bibitem[Spinella and Weber(2019)]{2019_fri}
Spinella, W.M.; Weber, F.
\newblock Hyperonic neutron star matter in light of GW170817.
\newblock {\em Astronomische Nachrichten} {\bf 2019}, {\em 340},~145–150.
\newblock {\url{https://doi.org/10.1002/asna.201913579}}.

\bibitem[Weber(2016)]{Weber2016}
Weber, F., Strange Quark Matter Inside Neutron Stars.
\newblock In {\em Handbook of Supernovae}; Alsabti, A.W.; Murdin, P., Eds.;
  Springer International Publishing: Cham,  2016; pp. 1--24.
\newblock {\url{https://doi.org/10.1007/978-3-319-20794-0_71-1}}.

\bibitem[Terazawa(2001)]{Terazawa:2001gg}
Terazawa, H.
\newblock {A new trend in high-energy physics: Current topics in nuclear and
  particle physics}.
\newblock In Proceedings of the {International Conference on New Trends in
  High-Energy Physics: Experiment, Phenomenology, Theory},  2001, pp. 246--255.

\bibitem[Husain and Thomas(2021)]{Husain_2021}
Husain, W.; Thomas, A.W.
\newblock Hybrid stars with hyperons and strange quark matter.
\newblock {\em PROCEEDINGS OF THE 14TH ASIA-PACIFIC PHYSICS CONFERENCE} {\bf
  2021}.
\newblock {\url{https://doi.org/10.1063/5.0036994}}.

\bibitem[Lattimer and Prakash(2001)]{2001_lattimer}
Lattimer, J.M.; Prakash, M.
\newblock Neutron Star Structure and the Equation of State.
\newblock {\em The Astrophysical Journal} {\bf 2001}, {\em 550},~426–442.
\newblock {\url{https://doi.org/10.1086/319702}}.

\bibitem[Zhao and Lattimer(2020)]{2020_latti}
Zhao, T.; Lattimer, J.M.
\newblock Quarkyonic matter equation of state in beta-equilibrium.
\newblock {\em Physical Review D} {\bf 2020}, {\em 102}.
\newblock {\url{https://doi.org/10.1103/physrevd.102.023021}}.

\bibitem[Drischler \em{et~al.}(2021)Drischler, Han, Lattimer, Prakash, Reddy,
  and Zhao]{2021_l}
Drischler, C.; Han, S.; Lattimer, J.M.; Prakash, M.; Reddy, S.; Zhao, T.
\newblock Limiting masses and radii of neutron stars and their implications.
\newblock {\em Physical Review C} {\bf 2021}, {\em 103}.
\newblock {\url{https://doi.org/10.1103/physrevc.103.045808}}.

\bibitem[Cierniak and Blaschke(2021)]{Cierniak:2021knt}
Cierniak, M.; Blaschke, D.
\newblock {Hybrid neutron stars in the mass-radius diagram}.
\newblock {\em Astron. Nachr.} {\bf 2021}, {\em 342},~819--825,
  \href{http://xxx.lanl.gov/abs/2106.06986}{{\normalfont
  [arXiv:nucl-th/2106.06986]}}.
\newblock {\url{https://doi.org/10.1002/asna.202114000}}.

\bibitem[Shahrbaf \em{et~al.}(2022)Shahrbaf, Blaschke, Typel, Farrar, and
  Alvarez-Castillo]{Shahrbaf:2022upc}
Shahrbaf, M.; Blaschke, D.; Typel, S.; Farrar, G.R.; Alvarez-Castillo, D.E.
\newblock {Sexaquark dilemma in neutron stars and its solution by quark
  deconfinement} {\bf 2022}.
\newblock  \href{http://xxx.lanl.gov/abs/2202.00652}{{\normalfont
  [arXiv:nucl-th/2202.00652]}}.

\bibitem[Nishizaki \em{et~al.}(2002)Nishizaki, Yamamoto, and
  Takatsuka]{10.1143/PTP.108.703}
Nishizaki, S.; Yamamoto, Y.; Takatsuka, T.
\newblock {Hyperon-Mixed Neutron Star Matter and Neutron Stars*)}.
\newblock {\em Progress of Theoretical Physics} {\bf 2002}, {\em
  108},~703--718,
  \href{http://xxx.lanl.gov/abs/https://academic.oup.com/ptp/article-pdf/108/4/703/5414579/108-4-703.pdf}{{\normalfont
  [https://academic.oup.com/ptp/article-pdf/108/4/703/5414579/108-4-703.pdf]}}.
\newblock {\url{https://doi.org/10.1143/PTP.108.703}}.

\bibitem[Yamamoto \em{et~al.}(2017)Yamamoto, Togashi, Tamagawa, Furumoto,
  Yasutake, and Rijken]{2017_xyz}
Yamamoto, Y.; Togashi, H.; Tamagawa, T.; Furumoto, T.; Yasutake, N.; Rijken,
  T.A.
\newblock Neutron-star radii based on realistic nuclear interactions.
\newblock {\em Physical Review C} {\bf 2017}, {\em 96}.
\newblock {\url{https://doi.org/10.1103/physrevc.96.065804}}.

\bibitem[Yamamoto \em{et~al.}(2022)Yamamoto, Yasutake, and Rijken]{2022_xxyz}
Yamamoto, Y.; Yasutake, N.; Rijken, T.A.
\newblock Quark-quark interaction and quark matter in neutron stars.
\newblock {\em Physical Review C} {\bf 2022}, {\em 105}.
\newblock {\url{https://doi.org/10.1103/physrevc.105.015804}}.

\bibitem[Motta and Thomas(2022)]{Motta:2022nlj}
Motta, T.F.; Thomas, A.W.
\newblock {The role of baryon structure in neutron stars}.
\newblock {\em Mod. Phys. Lett. A} {\bf 2022}, {\em 37},~2230001,
  \href{http://xxx.lanl.gov/abs/2201.11549}{{\normalfont
  [arXiv:nucl-th/2201.11549]}}.
\newblock {\url{https://doi.org/10.1142/S0217732322300014}}.

\bibitem[Guichon(1988)]{Guichon:1987jp}
Guichon, P.A.M.
\newblock {A Possible Quark Mechanism for the Saturation of Nuclear Matter}.
\newblock {\em Phys. Lett. B} {\bf 1988}, {\em 200},~235--240.
\newblock {\url{https://doi.org/10.1016/0370-2693(88)90762-9}}.

\bibitem[Guichon \em{et~al.}(1996)Guichon, Saito, Rodionov, and
  Thomas]{Guichon:1995ue}
Guichon, P.A.M.; Saito, K.; Rodionov, E.N.; Thomas, A.W.
\newblock {The Role of nucleon structure in finite nuclei}.
\newblock {\em Nucl. Phys. A} {\bf 1996}, {\em 601},~349--379,
  \href{http://xxx.lanl.gov/abs/nucl-th/9509034}{{\normalfont
  [nucl-th/9509034]}}.
\newblock {\url{https://doi.org/10.1016/0375-9474(96)00033-4}}.

\bibitem[Stone \em{et~al.}(2016)Stone, Guichon, Reinhard, and
  Thomas]{Stone:2016qmi}
Stone, J.R.; Guichon, P.A.M.; Reinhard, P.G.; Thomas, A.W.
\newblock {Finite Nuclei in the Quark-Meson Coupling Model}.
\newblock {\em Phys. Rev. Lett.} {\bf 2016}, {\em 116},~092501,
  \href{http://xxx.lanl.gov/abs/1601.08131}{{\normalfont
  [arXiv:nucl-th/1601.08131]}}.
\newblock {\url{https://doi.org/10.1103/PhysRevLett.116.092501}}.

\bibitem[{Rikovska Stone} \em{et~al.}(2007){Rikovska Stone}, Guichon,
  Matevosyan, and Thomas]{RIKOVSKASTONE2007341}
{Rikovska Stone}, J.; Guichon, P.; Matevosyan, H.; Thomas, A.
\newblock Cold uniform matter and neutron stars in the quark–meson-coupling
  model.
\newblock {\em Nuclear Physics A} {\bf 2007}, {\em 792},~341--369.
\newblock
  {\url{https://doi.org/https://doi.org/10.1016/j.nuclphysa.2007.05.011}}.

\bibitem[Saito \em{et~al.}(2007)Saito, Tsushima, and Thomas]{Saito:2005rv}
Saito, K.; Tsushima, K.; Thomas, A.W.
\newblock {Nucleon and hadron structure changes in the nuclear medium and
  impact on observables}.
\newblock {\em Prog. Part. Nucl. Phys.} {\bf 2007}, {\em 58},~1--167,
  \href{http://xxx.lanl.gov/abs/hep-ph/0506314}{{\normalfont
  [hep-ph/0506314]}}.
\newblock {\url{https://doi.org/10.1016/j.ppnp.2005.07.003}}.

\bibitem[DeGrand \em{et~al.}(1975)DeGrand, Jaffe, Johnson, and
  Kiskis]{DeGrand:1975cf}
DeGrand, T.A.; Jaffe, R.L.; Johnson, K.; Kiskis, J.E.
\newblock {Masses and Other Parameters of the Light Hadrons}.
\newblock {\em Phys. Rev. D} {\bf 1975}, {\em 12},~2060.
\newblock {\url{https://doi.org/10.1103/PhysRevD.12.2060}}.

\bibitem[Guichon \em{et~al.}(2018)Guichon, Stone, and Thomas]{Guichon:2018uew}
Guichon, P.A.M.; Stone, J.R.; Thomas, A.W.
\newblock {Quark\textendash{}Meson-Coupling (QMC) model for finite nuclei,
  nuclear matter and beyond}.
\newblock {\em Prog. Part. Nucl. Phys.} {\bf 2018}, {\em 100},~262--297,
  \href{http://xxx.lanl.gov/abs/1802.08368}{{\normalfont
  [arXiv:nucl-th/1802.08368]}}.
\newblock {\url{https://doi.org/10.1016/j.ppnp.2018.01.008}}.

\bibitem[Motta \em{et~al.}(2019)Motta, Kalaitzis, Anti\'c, Guichon, Stone, and
  Thomas]{Motta:2019tjc}
Motta, T.F.; Kalaitzis, A.M.; Anti\'c, S.; Guichon, P.A.M.; Stone, J.R.;
  Thomas, A.W.
\newblock {Isovector Effects in Neutron Stars, Radii and the GW170817
  Constraint}.
\newblock {\em Astrophys. J.} {\bf 2019}, {\em 878},~159,
  \href{http://xxx.lanl.gov/abs/1904.03794}{{\normalfont
  [arXiv:nucl-th/1904.03794]}}.
\newblock {\url{https://doi.org/10.3847/1538-4357/ab218e}}.

\bibitem[Krein \em{et~al.}(1999)Krein, Thomas, and Tsushima]{Krein:1998vc}
Krein, G.; Thomas, A.W.; Tsushima, K.
\newblock {Fock terms in the quark meson coupling model}.
\newblock {\em Nucl. Phys. A} {\bf 1999}, {\em 650},~313--325,
  \href{http://xxx.lanl.gov/abs/nucl-th/9810023}{{\normalfont
  [nucl-th/9810023]}}.
\newblock {\url{https://doi.org/10.1016/S0375-9474(99)00117-7}}.

\bibitem[Tolman(1934)]{Tolman169}
Tolman, R.C.
\newblock Effect of Inhomogeneity on Cosmological Models.
\newblock {\em Proceedings of the National Academy of Sciences} {\bf 1934},
  {\em 20},~169--176,
  \href{http://xxx.lanl.gov/abs/https://www.pnas.org/content/20/3/169.full.pdf}{{\normalfont
  [https://www.pnas.org/content/20/3/169.full.pdf]}}.
\newblock {\url{https://doi.org/10.1073/pnas.20.3.169}}.

\bibitem[Oppenheimer and Volkoff(1939)]{PhysRev.55.374}
Oppenheimer, J.R.; Volkoff, G.M.
\newblock On Massive Neutron Cores.
\newblock {\em Phys. Rev.} {\bf 1939}, {\em 55},~374--381.
\newblock {\url{https://doi.org/10.1103/PhysRev.55.374}}.

\bibitem[Hinderer(2008)]{Hinderer_2008}
Hinderer, T.
\newblock Tidal Love Numbers of Neutron Stars.
\newblock {\em The Astrophysical Journal} {\bf 2008}, {\em 677},~1216–1220.
\newblock {\url{https://doi.org/10.1086/533487}}.

\bibitem[Hinderer \em{et~al.}(2010)Hinderer, Lackey, Lang, and
  Read]{Hinderer_2010}
Hinderer, T.; Lackey, B.D.; Lang, R.N.; Read, J.S.
\newblock Tidal deformability of neutron stars with realistic equations of
  state and their gravitational wave signatures in binary inspiral.
\newblock {\em Physical Review D} {\bf 2010}, {\em 81}.
\newblock {\url{https://doi.org/10.1103/physrevd.81.123016}}.

\bibitem[Cline and Cornell(2018)]{Cline:2018ami}
Cline, J.M.; Cornell, J.M.
\newblock {Dark decay of the neutron}.
\newblock {\em JHEP} {\bf 2018}, {\em 07},~081,
  \href{http://xxx.lanl.gov/abs/1803.04961}{{\normalfont
  [arXiv:hep-ph/1803.04961]}}.
\newblock {\url{https://doi.org/10.1007/JHEP07(2018)081}}.

\bibitem[Berryman \em{et~al.}(2022)Berryman, Gardner, and
  Zakeri]{Berryman:2022zic}
Berryman, J.M.; Gardner, S.; Zakeri, M.
\newblock {Neutron Stars with Baryon Number Violation, Probing Dark Sectors}
  {\bf 2022}.
\newblock  \href{http://xxx.lanl.gov/abs/2201.02637}{{\normalfont
  [arXiv:hep-ph/2201.02637]}}.

\bibitem[Rajendran and Ramani(2021)]{2021_ramani}
Rajendran, S.; Ramani, H.
\newblock Composite solution to the neutron lifetime anomaly.
\newblock {\em Physical Review D} {\bf 2021}, {\em 103}.
\newblock {\url{https://doi.org/10.1103/physrevd.103.035014}}.

\bibitem[Tang \em{et~al.}(2018)Tang, Blatnik, Broussard, Choi, Clayton,
  Cude-Woods, Currie, Fellers, Fries, Geltenbort, Gonzalez, Hickerson, Ito,
  Liu, MacDonald, Makela, Morris, O’Shaughnessy, Pattie, Plaster, Salvat,
  Saunders, Wang, Young, and Zeck]{2018_tanga}
Tang, Z.; Blatnik, M.; Broussard, L.; Choi, J.; Clayton, S.; Cude-Woods, C.;
  Currie, S.; Fellers, D.; Fries, E.; Geltenbort, P.;  et~al.
\newblock Search for the Neutron Decay n$\rightarrow$ $\chi+\gamma$ , Where
  $\chi$ is a Dark Matter Particle.
\newblock {\em Physical Review Letters} {\bf 2018}, {\em 121}.
\newblock {\url{https://doi.org/10.1103/physrevlett.121.022505}}.

\bibitem[Berezhiani \em{et~al.}(2021)Berezhiani, Biondi, Mannarelli, and
  Tonelli]{2021_osc}
Berezhiani, Z.; Biondi, R.; Mannarelli, M.; Tonelli, F.
\newblock Neutron-mirror neutron mixing and neutron stars.
\newblock {\em The European Physical Journal C} {\bf 2021}, {\em 81}.
\newblock {\url{https://doi.org/10.1140/epjc/s10052-021-09806-1}}.

\bibitem[McKeen \em{et~al.}(2018)McKeen, Nelson, Reddy, and
  Zhou]{McKeen:2018xwc}
McKeen, D.; Nelson, A.E.; Reddy, S.; Zhou, D.
\newblock {Neutron stars exclude light dark baryons}.
\newblock {\em Phys. Rev. Lett.} {\bf 2018}, {\em 121},~061802,
  \href{http://xxx.lanl.gov/abs/1802.08244}{{\normalfont
  [arXiv:hep-ph/1802.08244]}}.
\newblock {\url{https://doi.org/10.1103/PhysRevLett.121.061802}}.

\bibitem[\"Ozel and Freire(2016)]{Ozel:2016oaf}
\"Ozel, F.; Freire, P.
\newblock {Masses, Radii, and the Equation of State of Neutron Stars}.
\newblock {\em Ann. Rev. Astron. Astrophys.} {\bf 2016}, {\em 54},~401--440,
  \href{http://xxx.lanl.gov/abs/1603.02698}{{\normalfont
  [arXiv:astro-ph.HE/1603.02698]}}.
\newblock {\url{https://doi.org/10.1146/annurev-astro-081915-023322}}.

\bibitem[Bramante \em{et~al.}(2018)Bramante, Linden, and
  Tsai]{Bramante_2018_31}
Bramante, J.; Linden, T.; Tsai, Y.D.
\newblock Searching for dark matter with neutron star mergers and quiet
  kilonovae.
\newblock {\em Phys. Rev. D} {\bf 2018}, {\em 97},~055016.
\newblock {\url{https://doi.org/10.1103/PhysRevD.97.055016}}.

\bibitem[Antypas \em{et~al.}(2022)Antypas et~al.]{Antypas:2022asj}
Antypas, D.;  et~al.
\newblock {New Horizons: Scalar and Vector Ultralight Dark Matter} {\bf 2022}.
\newblock  \href{http://xxx.lanl.gov/abs/2203.14915}{{\normalfont
  [arXiv:hep-ex/2203.14915]}}.

\bibitem[Cadamuro and Redondo(2012)]{Cadamuro:2011fd}
Cadamuro, D.; Redondo, J.
\newblock {Cosmological bounds on pseudo Nambu-Goldstone bosons}.
\newblock {\em JCAP} {\bf 2012}, {\em 02},~032,
  \href{http://xxx.lanl.gov/abs/1110.2895}{{\normalfont
  [arXiv:hep-ph/1110.2895]}}.
\newblock {\url{https://doi.org/10.1088/1475-7516/2012/02/032}}.

\bibitem[Caputo \em{et~al.}(2021)Caputo, O'Hare, Millar, and
  Vitagliano]{Caputo:2021eaa}
Caputo, A.; O'Hare, C.A.J.; Millar, A.J.; Vitagliano, E.
\newblock {Dark photon limits: a cookbook} {\bf 2021}.
\newblock  \href{http://xxx.lanl.gov/abs/2105.04565}{{\normalfont
  [arXiv:hep-ph/2105.04565]}}.

\bibitem[Aprile \em{et~al.}(2022)Aprile, Abe, Agostini, Ahmed~Maouloud,
  Alfonsi, Althueser, Angelino, Angevaare, Antochi, Ant\'on~Martin, Arneodo,
  Baudis, Baxter, Bellagamba, Bernard, Biondi, Bismark, Brown, Bruenner, Bruno,
  Budnik, Capelli, Cardoso, Cichon, Cimmino, Clark, Colijn, Conrad,
  Cuenca-Garc\'{\i}a, Cussonneau, D'Andrea, Decowski, Di~Gangi, Di~Pede,
  Di~Giovanni, Di~Stefano, Diglio, Elykov, Farrell, Ferella, Fischer, Fulgione,
  Gaemers, Gaior, Galloway, Gao, Glade-Beucke, Grandi, Grigat, Higuera, Hils,
  Hoetzsch, Howlett, Iacovacci, Itow, Jakob, Joerg, Joy, Kato, Kavrigin,
  Kazama, Kobayashi, Koltman, Kopec, Landsman, Lang, Levinson, Li, Li, Liang,
  Lindemann, Lindner, Liu, Lombardi, Long, Lopes, Ma, Macolino, Mahlstedt,
  Mancuso, Manenti, Manfredini, Marignetti, Marrod\'an~Undagoitia, Martens,
  Masbou, Masson, Masson, Mastroianni, Messina, Miuchi, Mizukoshi, Molinario,
  Moriyama, Mor\aa{}, Mosbacher, Murra, M\"uller, Ni, Oberlack, Paetsch,
  Palacio, Peres, Pienaar, Pierre, Pizzella, Plante, Qi, Qin,
  Ram\'{\i}rez~Garc\'{\i}a, Reichard, Rocchetti, Rupp, Sanchez, dos Santos,
  Sarnoff, Sartorelli, Schreiner, Schulte, Schulze~Ei\ss{}ing, Schumann,
  Scotto~Lavina, Selvi, Semeria, Shagin, Shi, Shockley, Silva, Simgen, Takeda,
  Tan, Terliuk, Thers, Toschi, Trinchero, Tunnell, T\"onnies, Valerius, Volta,
  Wei, Weinheimer, Weiss, Wenz, Wittweg, Wolf, Xu, Yamashita, Yang, Ye, Yuan,
  Zavattini, Zhang, Zhong, Zhu, and Zopounidis]{PhysRevD.106.022001}
Aprile, E.; Abe, K.; Agostini, F.; Ahmed~Maouloud, S.; Alfonsi, M.; Althueser,
  L.; Angelino, E.; Angevaare, J.R.; Antochi, V.C.; Ant\'on~Martin, D.;  et~al.
\newblock Emission of single and few electrons in XENON1T and limits on light
  dark matter.
\newblock {\em Phys. Rev. D} {\bf 2022}, {\em 106},~022001.
\newblock {\url{https://doi.org/10.1103/PhysRevD.106.022001}}.

\end{thebibliography}



\end{document}